# Evidence of diamagnetic fluctuations above $T_c$ in BaFe$_{2-x}$Co$_x$As$_2$ single crystals

Biplab Bag[1], K. Vinod[2], A. Bharathi[2] and S. S. Banerjee[1*]

[1]Department of Physics, Indian Institute of Technology, Kanpur-208016, India

[2]Condensed Matter Physics Division, Materials Science Group, Indira Gandhi Center for Atomic Research, Kalpakkam-603102, India

*Email: satyajit@iitk.ac.in

**Abstract**

**Increasing doping concentration ($x$) in Pnictide superconductors results in the suppression of long range magnetic order and the emergence of superconductivity. While doping destroys long range magnetic order however the effect if any of the surviving magnetic fluctuations on superconductivity in Pnictides is currently not well understood. In optimally doped BaFe$_{2-x}$Co$_x$As$_2$ single crystals, below $T_c$ we observe local regions with positive magnetization coexisting along with superconductivity at low applied magnetic fields ($H$). With increasing $H$ the positive magnetization response gets weaker and the robust superconducting diamagnetic response appears. The normal state is found to be inhomogeneous with regions having local diamagnetic response embedded in a positive magnetization background. Estimates of the superconducting fraction in the normal state shows it maximizes for the optimally doped crystal. We construct a doping dependent $H$ - $T$ diagram identifying different fluctuation regimes and our data suggests a close correlation between magnetic and superconducting fluctuations in BaFe$_{2-x}$Co$_x$As$_2$.**

**Introduction**

The doping phase diagram of Pnictide superconductors shows magneto-structural transformation boundaries interrupted by a superconducting dome [1,2,3,4,5,6,7,8]. Unraveling the intricate relationship between superconductivity, magnetism and the structural changes in Pnictide superconductors is a topic of ongoing interest. Doping dependence studies in Pnictides indicate that superconductivity arises following destruction of long range magnetic order, with the bulk superconducting transition temperature ($T_c$) becoming maximum at an optimal doping concentration ($x$). MuSR studies on Ba(Fe$_{1-x}$Co$_x$)$_2$As$_2$ have shown that Co doping results in loss of long range magnetic order with a disordered magnetic or incommensurate spin density wave state [9,10,11,12] being present. Recent Iron isotope variation studies in doped 1111 and 122 Pnictides show that both the magnetic ($T_N$) and bulk superconducting transition temperatures ($T_c$) are affected, suggesting the possibility of a close relationship between magnetism and superconductivity [13] in these compounds. Recent studies at temperatures ($T$) below $T_c$ in underdoped Pnictides suggest the coexistence of superconductivity (SC) along with magnetic fluctuations [14,15,16,17,18,19,20]. Theories suggest that magnetism plays an important role in mediating strong superconducting pairing correlations in Pnictides [21,22,23,24,25,26,27,28,29]. Thus while doping dependence studies suggest that suppression of long range magnetic order in Pnictides is important for the emergence of superconductivity one needs to reconcile with evidences that magnetism is also important for superconductivity in these compounds. Here we consider following: if magnetic fluctuations are important for mediating superconducting pairing and the fluctuations are strong enough to survive above $T_c$ then it is plausible that they may mediate local superconducting fluctuations to exist in the normal state. Evidence for the presence of superconducting fluctuations above $T_c$ in Pnictides isn't a well resolved issue. ARPES study in a BaFe$_2$(As$_{1-x}$P$_x$)$_2$ indicates the presence of local superconducting correlations above bulk $T_c$ [30]. However in Ba(Fe$_{1-x}$Co$_x$)$_2$As$_2$ system STM study on crystals with different $x$, suggests the superconducting gap closes at $T_c$ [31]. Recently there is a study in FeSe crystal which shows evidence for the presence of preformed pairs above $T_c$ [32]. In this article we present sensitive measurements of local and bulk magnetization in BaFe$_{2-x}$Co$_x$As$_2$ single crystals with the Co concentration ($x$) varied from under to over doped regime. All the crystals show a robust bulk diamagnetic Meissner shielding response at moderate applied magnetic field ($H$). At $T < T_c$, local magnetic field measurements reveal the presence of a weak positive magnetization (paramagnetic) response in the optimally doped crystal at very low applied $H$ which transforms into a diamagnetic response with increase in $H$. In all the crystals (with different $x$) just above $T_c$ we observe positive magnetization response which shows a peak like feature and a long paramagnetic tail extending upto high $T$. Analysis of this feature as well as imaging of differential changes in local magnetic field in response to external $H$ modulation reveal the presence of an inhomogeneous normal state with local regions with enhanced diamagnetic shielding response embedded in a background with positive magnetization response. We propose the presence of local superconducting fluctuations exhibiting with local diamagnetic response in the normal state of this compound. Above $T_c$, we investigate the variation of superconducting volume fraction with $x$. We observe that the crystal with optimum doping concentration exhibits the strongest effects of magnetic fluctuations and it also possesses the highest volume fraction with superconducting fluctuations above $T_c$. Using the data gathered from each crystal we identify different characteristic temperatures measured as a function of $H$, which we summarize in a $H$ - $T$ diagram for each $x$. Using this diagram we identify different regimes, viz., a regime where exclusive magnetic fluctuations exist, a regime above $T_c$ where superconducting and magnetic fluctuations coexist and a regime below $T_c$ where weak magnetic fluctuations coexist with bulk superconductivity. We believe our results suggests the presence of local superconducting fluctuations above $T_c$ and that magnetic fluctuations play a role in mediating superconducting order in the BaFe$_{2-x}$Co$_x$As$_2$ system.

Single crystals of BaFe$_{2-x}$Co$_x$As$_2$ with $x$ = 0.10, 0.14, and 0.20 were prepared using self-flux method [33]. The bulk $T_c$ is identified from the onset of diamagnetism in $T$ dependent magnetization $M(T)$ at low $H$ and from the sharp drop in resistance $R(T)$ (c.f. arrows in Figure 1a-c). The $T_c$'s determined from the two measurements differ slightly (see Table 1) as $T_c$ determined from bulk $M(T)$ measurements is with a low $H$ = 50 Oe while $T_c$ determined from $R(T)$ is at $H$ = 0 Oe, also the $M(T)$ and $R(T)$ measurements were performed in different cryogenic systems. The variation of $T_c$ with Co doping concentration ($x$) is summarized in Table 1. Using the $R(T)$ values, we label the crystal $x$ = 0.14 (Co-14) as being close to Optimal doping (OpD) (highest $T_c(0)$ = 26.80 K), Co-10 ($x$ = 0.10) as

**Table 1**

| Crystal | $x_{nominal}$ | $x_{EPMA}$ | $T_c$ (K)* | Volume $l \times w \times d$ (mm$^3$) |
|---|---|---|---|---|
| Co-10 (Under-doped (UD)) | 0.10 | 0.1032 ± 0.0003 | 18.10 [17.04] ± 0.05 | 0.93 x 0.83 x 0.13 |
| Co-14 (Optimally-doped (OpD)) | 0.14 | 0.1266 ± 0.0002 | 26.80 [25.47] ± 0.05 | 0.61 x 0.26 x 0.05 |
| Co-20 (Over-doped (OD)) | 0.20 | 0.1986 ± 0.0002 | 25.99 [24.98] ± 0.05 | 1.62 x 0.69 x 0.17 |

*Estimated from $R(T)$ at $H$=0 Oe[$M(T)$ at 50 Oe].

underdoped (UD) and Co-20 ($x$ = 0.20) as overdoped (OD) crystal, consistent with values in earlier reports [1] (note in ref. 1 the $x$ is defined as in Ba(Fe$_{1-x}$Co$_x$)$_2$As$_2$ while in our article we define $x$ as BaFe$_{2-x}$Co$_x$As$_2$). The RRR for Co-10, Co-14 and Co-20 crystals were 1.92, 2.45 and 2.71, respectively, and resistivity at 30 K in the range of 100 $\mu\Omega$-cm. In all our measurements the applied $H$ is parallel to [001] direction of the single crystal. The local magnetic field distribution ($B_z(x,y)$) at different $H$ and $T$ across the sample surface is mapped using high sensitivity magneto-optical (MO) imaging [34,35,36] technique (see supplementary).

Inset of Figure 1a shows the normalized $R(T)$ (normalized w.r.t $R$ at 300 K) behavior for all three crystals at $H$ = 0 Oe. The normalized $R(T)$ for Co-10 (UD) crystal shows an anomalous upturn in $R$ at 67 K, which is a feature associated with magneto-structural transition in Pnictides (c.f. ref. 1: similar features in Ba(Fe$_{1-x}$Co$_x$)$_2$As$_2$ with $x$ = 0.05). For the OpD and OD crystals, the absence of any such anomaly in $R(T)$ suggests the suppression of long range magnetic order with increase in Co doping. Using bulk $M(T)$ measurements at different $H$, Figure 1a-c show the behavior of the bulk zero field cooled (ZFC) $4\pi M / H$ as a function of $T$ at different $H$ for Co-10, Co-14 and

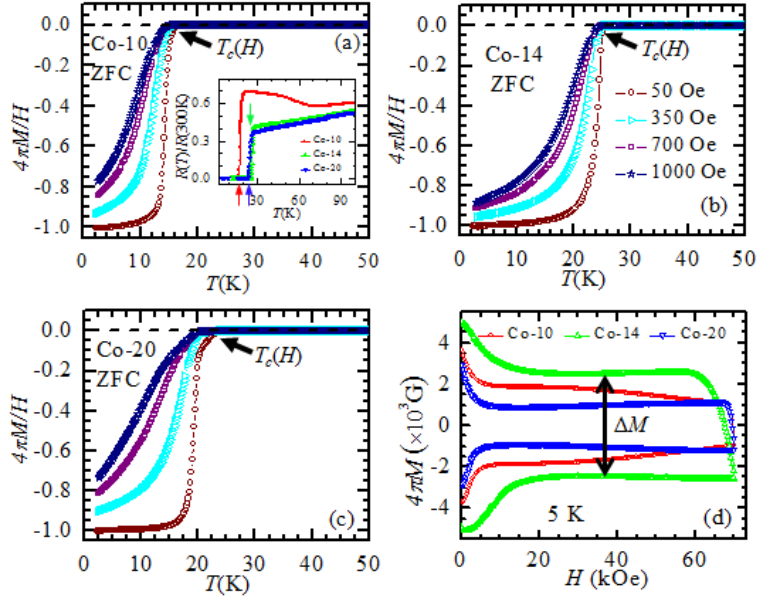

**Figure 1** Variation of *4πM/H* (bulk DC magnetic susceptibility with demagnetization correction) with temperature (ZFC warming) for (a) Co-10, (b) Co-14 and (c) Co-20 respectively at different fields applied parallel to [001] direction (along c-axis) of the crystal. The onset of diamagnetic signal at $T_c(H)$ is identified by arrows for the three crystals. The field legends are identical across (a)-(c). Inset of (a) shows the temperature dependent electrical resistance of Co-10, Co-14 and Co-20 crystals normalized to their room temperature resistance at *H*=0 Oe. (d) Bulk DC magnetization hysteresis loop (forward and reverse legs corresponds to 5[th] and 6[th] quadrant legs respectively of a 6 quadrant *M(H)* hysteresis loop) for Co-10, Co-14 and Co-20 crystals as a function of *H* at 5 K. The width of the magnetic hysteresis loop Δ*M(H)* at *H* for Co-14 is shown with the arrow.

Co-20 crystals respectively. At low *T* (<<$T_c$) and low *H* the $4\pi M/H$ saturates to a value close to -1 which is expected from a Meissner state of the superconductor. Figure 1a-c show that at a fixed *T* the diamagnetic $4\pi M/H$ value decreases from -1 with increase in *H*, due to the penetration of vortices into the superconductor. Figure 1d shows conventional superconducting hysteresis *M(H)* loops measured at 5 K (< $T_c$) for UD (red), OpD (green) and OD (blue) crystals. One may note that, the above features in *M(T)* and *M(H)* in Figures 1a-d for all the three crystals show bulk pinning and robust bulk type II superconductivity in all the three single crystals. In Figure 1d the width of the hysteresis loop Δ*M* gives an estimate of the critical current density, $J_c = 20\Delta M / a(1-a/3b)$ [37] (for a rectangular plate like crystal geometry, with *a*, *b* (with b > a) as the in-plane crystal dimensions) in these superconductors. The maximum $J_c$ for our OpD crystal is ~ $8\times10^5$ Amp/cm$^2$ which is comparable to $J_c$'s found in these superconductors [38].

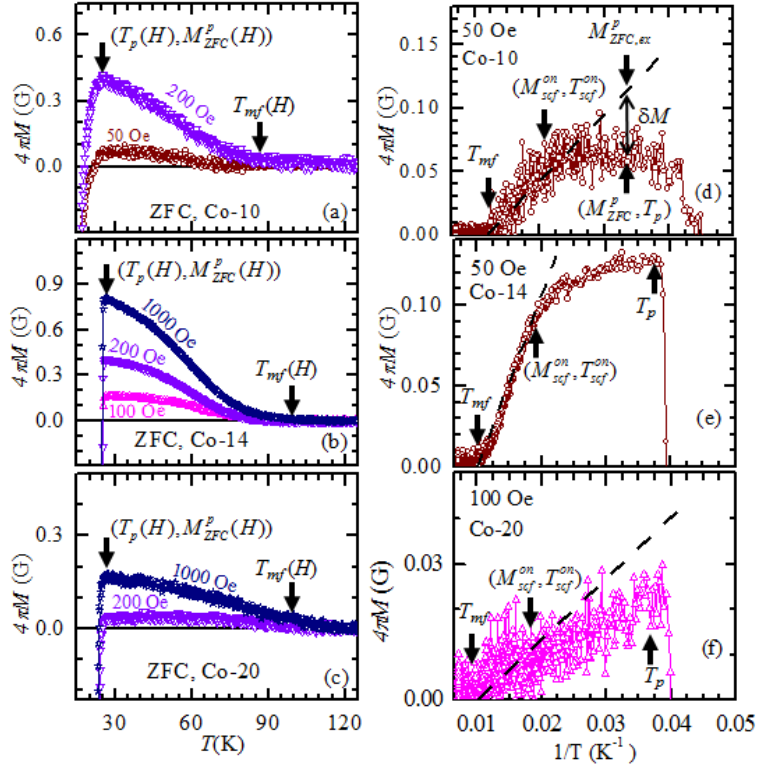

**Figure 2** Zoomed-in view of ZFC $4\pi M/H$ vs $T$ above $T_c$ for (a) Co-10, (b) Co-14 and (c) Co-20 respectively at different fields. The onset of magnetic fluctuation ($T_{mf}^{on}$) and the location of the peak of the positive magnetization response ($T_p$ and $M_{ZFC}^{p}$) have been identified with arrows for the three samples. Isofield ZFC $4\pi M$ vs $1/T$ at $T>T_c$ for (d) Co-10 (at $H$=50 Oe), (e) Co-14 (at $H$=50 Oe) and (f) Co-20 (at $H$=100 Oe) crystal respectively. The solid dashed lines represent linear fitting signifying $M \propto \frac{1}{T}$ behavior for the $T$-dependent magnetization (at $T>T_c$) for the three samples. The deviations of linear behavior from the experimental data have been identified using arrows at $T_{scf}^{on}$ which correspond to the onset of superconducting fluctuation. $T_{mf}^{on}$ and $T_p$ have been identified with arrows. The linear extrapolation of $4\pi M(1/T)$ at $T_p$ has been identified as $4\pi M_{ZFC,ex}^{p}$ and the quantity $\delta M = 4\pi M_{ZFC}^{p} - 4\pi M_{ZFC,ex}^{p}$, which corresponds the suppression of positive magnetization response due to presence of superconducting fluctuation at $T_p$, is marked with arrow in (d).

Figure 2a-c show the behavior of $M(T)$ near $T_c(H)$ and above it. While in any conventional superconductor, the diamagnetic $M(T)$ response is expected to monotonically approach zero at $T_c(H)$, instead the diamagnetic $M(T)$ in Figure 2a-c at $T < T_c(H)$ monotonically crosses zero and becomes positive (paramagnetic) above $T_c(H)$. The paramagnetic $M(T)$ response reaches a maximum value at $T_p$ with a peak height $M_{ZFC}^{p}$, beyond which $M$ decreases over a long paramagnetic tail extending upto $T >> T_c(H)$. At first glance the shape of the paramagnetic peak at $T_p$ can be mistaken for the Wohlleben or Paramagnetic Meissner Effect (PME) [39,40] where this effect is found only in Field Cooled (FC) $M(T)$ measurements. Non uniform cooling of a superconductor [39] in a field or surface superconductivity effects [40] generates flux compression leading to a positive PME peak near $T_c$ in a FC $M(T)$

measurements [39]. However note that the peak like feature in *M(T)* in Figure 2a-c have no similarity with PME because (i) the peak in Figure 2a-c is found in a ZFC *M(T)* measurement unlike PME which is found only in a FC measurement and (ii) the $M^p_{ZFC}$ increases with increasing *H* (c.f. Fig. 2a-c), unlike in PME where the paramagnetic peak in *M(T)* is found only at very low *H* and the peak height decreases with increasing *H* [39,40]. Infact our observation that $M^p_{ZFC}$ increases with *H* suggests a magnetic origin of this positive magnetic response found above $T_c$. The peak in *M(T)* above $T_c$ for Co-10 and Co-14 crystals are almost comparable in magnitude at similar values of *H* while it is comparatively weaker for the Co-20 crystal, although Co-20 has the highest Co doping concentration.

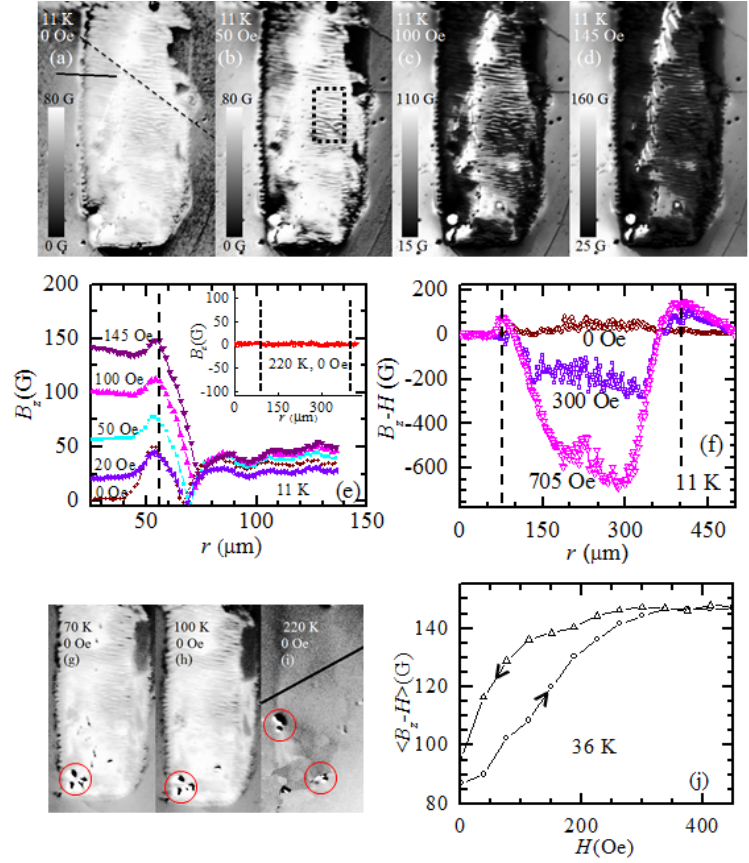

**Figure 3** Iso-thermal MO images (supplementary) for Co-14 at 11 K ($<T_c$) at (a) 0 Oe, (b) 50 Oe, (c) 100 Oe and (d) 145 Oe respectively with H//c. The color bars in inset shows the mapping of contrast to local field $B_z$. (e) The local $B_z(r)$ profile at different fields at 11 K measured along the continuous horizontal line shown in (a). The edge of the sample is identified with the vertical dashed line. The inset of (e) presents $B_z(r)$ distribution at 220 K and 0 Oe measured along the line shown in (i). (f) The variation of local magnetization ($B_z(r)$-H) for Co-14 at 11 K and different fields measured along the dashed line shown in (a). The dashed lines in (f) represent the edge of the sample. (g)-(i) Zero field MO images for Co-14 crystal at 70 K, 100 K and 220 K respectively. The red circles indicate scratches / defects on the MO indicator (see Supplementary). (j) The field dependence of $<B_z-H>$ (averaged over 750 µm$^2$ area) hysteresis loop (virgin forward and reverse as shown by arrows) for Co-14 at 36 K ($>T_c$) with H//c.

Figure 3a is a MO image (see Supplementary) of the local $B_z(x, y)$ distribution measured across the surface of Co-14 (OpD) crystal at 11 K ($<< T_c(0) = 26.80$ K) and $H = 0$ Oe. The image shows an unconventional feature, viz., at $H = 0$ Oe for a ZFC superconductor, the image contrast in Figure 3a inside the crystal is bright which implies a non-zero mean $B_z \sim 35$ G (c.f. Fig. 3e-f). This is unlike the dark magneto-optical (MO) contrast (see Supplementary) found in any conventional superconductor prepared in a ZFC Meissner state at low $H$, as $B_z \sim 0$ G. As $H$ is increased, the contrast inside the crystal turns from bright to dark (c.f. Fig. 3a-d) unlike a conventional superconductor which goes the other way as it enters the mixed state with $B_z \neq 0$ G with increasing $H$ from a Meissner state with $B_z = 0$ G. The darkening of the crystal (Fig. 3a-d) suggests an increase in diamagnetic shielding response with increase in $H$.

The $B_z(r)$ profile measured at different $H$ along a representative horizontal solid dark line (dashed line), where $r$ is the distance measured along the line (c.f. Fig. 3a), are plotted in Figure 3e (Figure 3f). Figure 3e shows that at $H = 0$ Oe deep inside the crystal the local $B_z$ is nonzero and has a finite value of about 35 G. Outside the crystal away from the crystal edges (identified by a vertical dashed line in Fig. 3e) $B_z \sim H$ and it increases with $H$. Near the crystal edge, $B_z$ is enhanced as there are diamagnetic shielding currents circulating around the edges of the superconductor. With increasing $H$ deep inside the crystal $B_z$ is found to change only by a small amount, for example, at $H=145$ Oe, $B_z \sim 40$ G which is close to the value observed at $H = 0$ Oe. The behavior of the local z - component of the magnetization of the sample is estimated using $B_z(r)$-$H$. From Figure 3e,f we note that for low $H$ in the range $0 \leq H \leq 40$ Oe, $B_z$-$H$ is positive (c.f. Fig. 5a shows $B_z$-$H$ for $H = 0$ Oe on expanded scale for Co-14 crystal), which turns into negative values (a characteristic feature of the diamagnetic response of superconductors) as $H$ is increased upto 705 Oe. This unusual transformation from a weakly positive magnetization at low $H$ into a strong diamagnetic response at moderately high $H$ will be discussed later.

It is to be noted that at 0 Oe applied field we observe a bright MOI contrast from the crystal sustained upto $T$ which is well above $T_c$ albeit with decreasing intensity (c.f. Fig. 3g-i) upto 220 K (Fig. 3i). Inset of Figure 3e shows a nearly featureless $B_z(r)$ distribution measured across the MO image at 220K and 0 Oe (Fig. 3i), indicating the positive $M(T)$ response from the crystal is quite weak at 220 K ($>> T_c$). The observation of positive $B_z$ at $H = 0$ Oe above $T_c$ (c.f. Fig. 3g-i) is attributed to the presence of regions with short range magnetic order in the Co-14 crystal. Earlier studies have suggested that doping destroys long range magnetic correlations but preserves locally short range magnetic order. Infact $\mu$SR and Neutron diffraction studies on Co doped $BaFe_2As_2$ system suggest that doping produces a disordered Fe magnetic configuration along with some magnetic order [9-11] preserved locally. In our measurements, the weak positive magnetic response emanating from regions with short range magnetic order shows following features: (i) paramagnetic $M(T)$ response above $T_c$ increases with $H$, (ii) we argue that feature (i) is not that a conventional paramagnet as, Figure 3j shows a hysteretic nature of the average $\langle B_z - H \rangle$ versus $H$ measured in the normal state of the superconductor at 36 K ($>T_c(0) = 26.80$ K). Furthermore the deviation of $\langle B_z - H \rangle$ from a linear $H$ dependent behavior in Figure 3j and the presence of a weak hysteresis suggest the response of the magnetic state above $T_c$ is unlike that of an ordinary paramagnet. The observation of a local positive $B_z$-$H$ at low $H$ (Fig. 3f) and a diamagnetic bulk $M(T)$ response below $T_c$ suggest the coexistence of short range magnetic order locally along with bulk superconductivity below $T_c$. While from the doping phase diagram of Pnictides coexistence of magnetic fluctuations along with superconductivity in underdoped $BaFe_{2-x}Co_xAs_2$ [9,10] is not a surprise, the sustenance of this coexistence to optimally doped regime is surprising. The Electron Probe Micro Analysis (EPMA) characterization of the crystals shows negligible local variation of Co concentration across the crystal, indicated by small error in the $x_{EPMA}$ value in Table 1. We would like to mention that from the MO images, positive $B_z$-$H$ response at low $H$ is found uniformly across the entire crystal (c.f. Fig. 3a, g-i) and it cannot be attributed to Co concentration variations across the crystal (later on we provide further evidence to discount this possibility further).

Based on the observations in Figure 2 and 3, we attribute the positive magnetization response above $T_c$ to set in from regions with short range magnetic order at $T \leq T_{mf}^{on}$ (*mf*: magnetic fluctuations). In the normal state of the crystals, Figures 2d-f show that the $M$ exhibits a linear paramagnetic Curie like behavior in $M$ vs $1/T$ extending upto a characteristic temperature $T_{scf}^{on}$. From Figure 2d-f we identify $T_{mf}^{on}$ as the onset $T$ of linear Curie like behavior in $M$. From our earlier observation of hysteresis and non-linear field dependence in the normal state (c.f. Fig. 3j), the Curie like behavior observed below $T_{mf}^{on}$ ($> T_c$) cannot be attributed to conventional paramagnetic response. We believe that due to the absence of a significant magnetic anisotropy energy scale in the system these local regions with magnetic correlations would be fluctuating about exhibiting a Curie like behavior at high $T$. Figure 6a-c show the $T_{mf}^{on}(H)$ line for Co-10, Co-14 and Co-20 crystals respectively. In Figure 2d-f, as $M$ approaches $T = T_{scf}^{on}$ we observe a large deviation of $M$ from the linear Curie behavior. We attribute this large deviation in $M$ below $T_{scf}^{on}$ to the onset of superconducting fluctuations (note $T_{scf}^{on} > T_c$) where weak diamagnetism arising from superconducting fluctuations suppresses the paramagnetic response emanating from regions with short range magnetic order. The $T_{scf}^{on}$ is found to be only weakly $H$ dependent.

From isofield MO images captured at different $T$ and $H$ = 200 Oe with ZFC warming for Co -14 crystal we determine the local $B_z(x, y)$ and we plot $\langle B_z - H \rangle$ vs $T$ in Figure 4a, where the average value <..> is obtained by averaging $(B_z(x, y) - H)$ over two regions marked S and B (c.f. inset Fig. 4b). Figure 4a shows peak like feature $\langle B_z(T) - H \rangle$ with a long decreasing positive magnetization tail in the S region, similar to 200 Oe bulk $4\pi M(T)$

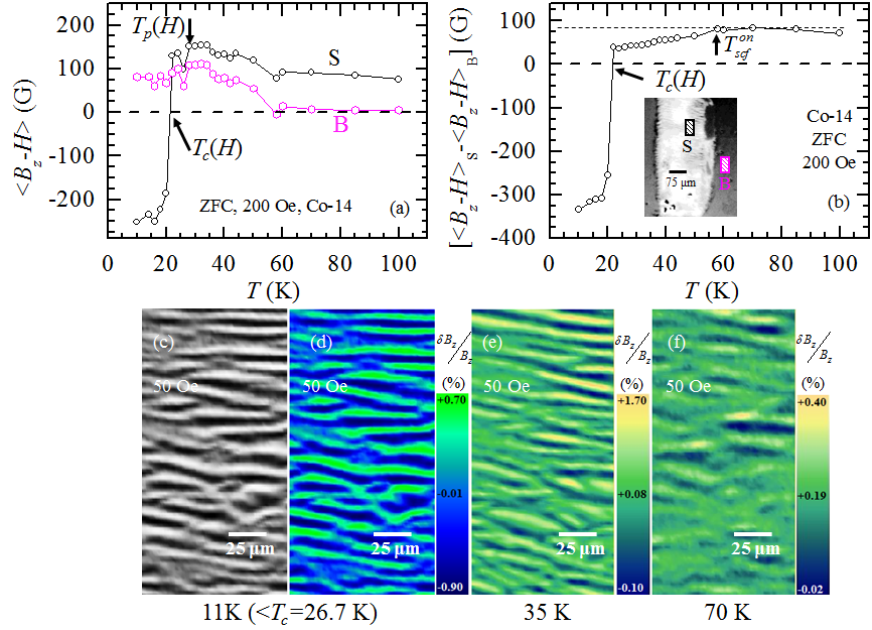

**Figure 4** (a) $\langle B_z\text{-}H\rangle$ vs $T$ inside (S) and outside (B) the crystal for ZFC state at 200 Oe obtained from MO imaging. Mean of ($B_z$-$H$) is obtained by averaging the signal over an area of 2000 μm² (areas marked in Inset of (b)). The B curve corresponds to the background data close to the sample edge. The position of $T_p$ and $T_c(H)$ have been marked with arrows. (b) The difference between $\langle B_z\text{-}H\rangle$ curves determined at S and B showing onset of superconducting fluctuation at $T_{scf}^{on}$ (marked with arrows). Inset presents a representative MO image at $H$=200 Oe and $T$=100 K showing the position of S and B regions. The dashed horizontal line around 100 G shows the offset of mean data. (c) Zoomed-in view of a portion (region marked in Fig. 3(b)) of a DMO image for Co-14 crystal at 50 Oe and 11 K. (d) The colored image of (c). The image is colored using the percentage value of $\delta B_z/B_z$ (see text for details). Zoomed colored DMO images for Co-14 at (e) 35K and (f) 70K respectively with $H$= 50 Oe.

data in Figure 2b. The sharp drop in magnetization due to onset of bulk superconductivity at $T_c(H)$ is also clearly discernible in Figure 4a (curve S). We have already shown in Figure 3g-i that the positive local $B_z$ due to magnetic fluctuations (or regions with short range magnetic order) surviving upto $T > T_c$. Above $T_c$, in a region B just outside the crystal the $\langle B_z - H \rangle$ value is determined by the positive magnetization due to the presence of a magnetic fraction inside the crystal. Therefore above $T_c$, from the response at S in Figure 4a we subtract the $\langle B_z(T) - H \rangle$ value determined in a region B to reduce magnetic contribution. While the temperature dependence of $\langle B_z - H \rangle$ at B and S are similar (c.f. Fig. 4a), there values are offset as the locations of B and S are different. The $\langle B_z - H \rangle$ value at B is expected to be slightly lower than that at S as seen in Figure 4a. Figure 4b shows the difference between the $\langle B_z(T) - H \rangle$ curves determined at S and B. Note that below the arrow marked $T_{scf}^{on}$ ~ 57 K, the drop in the $\langle B_z(T) - H \rangle$ below is clearly identifiable. From this analysis it is clear that the peak feature at $T_p$ in Figure 4a cannot be attributed to magnetic fluctuations alone. In Figure 4b by subtracting the mean offset between S and B (viz., the dashed horizontal line) such that the data becomes negative at $T < T_{scf}^{on} \sim 57$ K, suggests the presence of a weak diamagnetic tail between $T_{scf}^{on}(H)$ and $T_c(H)$ followed by a sharp transition into the bulk superconducting state below $T_c(H)$.

Figure 4c-f show images of Co-14 captured using a modified differential magneto-optical imaging technique (DMO) [41,42,43] [see Supplementary] obtained from a portion of the crystal (viz., region within the dashed black rectangular portion in Fig. 3b) at 50 Oe and at different $T$. The contrast in differential MO image represents changes in $B_z$, viz., $\delta B_z$, produced inside the crystal in response to modulation in external $H$ by $\delta H = 1$ Oe [see Supplementary]. The DMO image in Figure 4c shows bright and dark stripes at 11 K. From the calibrated DMO images we calculate the $\delta B_z / B_z(H)$ values at each pixel in the image. The $\delta B_z / B_z(H)$ values in percentage are used to color the gray images in Figure 4d-f with the color scale represented beside each image. Regions with dark and lighter shade of blue in Figure 4d correspond to superconducting regions at 11 K with negative $\delta B_z$. The maximum negative value of $\delta B_z / B_z(H) \sim -0.90$ % is over darkest blue region. From Figure 4d the inhomogeneous nature of the superconducting state below $T_c$ is clear. We use negative $\delta B_z / B_z(H)$ values to identify superconducting regions below $T_c$ at low $H$. The positive $\delta B_z / B_z(H)$ values (greenish-yellowish shaded) in Figure 4d are found in regions with short range magnetic order (regions with magnetic fluctuations). The optimally doped Co-14 crystal below $T_c$ (Fig. 4d) shows a state with predominating superconducting region coexisting with magnetic regions localized over greenish stripe like regions. Figure 4e-f show that above $T_c$ most of the regions within the crystal are positively magnetized due to predominance of magnetic fluctuations however locally there are regions with a weak diamagnetic response viz., negative $\delta B_z / B_z(H)$ response. Thus the normal state is also inhomogeneous and exhibits regions with local diamagnetic response embedded in a background with predominant positive magnetization. It may be worth mentioning that an earlier study had observed below $T_c$, the presence of stripe like regions with enhanced diamagnetic shielding response located over twin boundaries in underdoped $BaFe_{2-x}Co_xAs_2$ crystals [44]. The regions with weak diamagnetic response we observe in our optimally doped crystal represent superconducting fluctuations above $T_c$ which survive until $T = T_{scf}^{on}$. This feature is consistent with the observation of a weak diamagnetic tail extending upto 57 K in Figure 4b. We observe that by 70 K the density of the bluish regions have reduced significantly. Based on the above we also understand that the deviation of $M$ from the Curie like behavior in Figure 2d-f, is due to the onset of the onset of diamagnetic fluctuations below $T_{scf}^{on}$. The relatively large size of the local regions with diamagnetic fluctuations above $T_c$ could be a result of the spatial resolution limit of our technique due to which are unable to resolve microscopic features. It is also possible these regions with local diamagnetism above $T_c$ are nucleated around extended defect in these crystals [44].

In Figure 5 the superconducting and magnetic response at $T < T_c$ is compared for Co-14 (OpD) and Co-20 (OD) crystals. A comparison of the local $B_z$ distribution in Co-14 and Co-20 crystals measured at $T$ =11 K ($<T_c$) and $H$ =

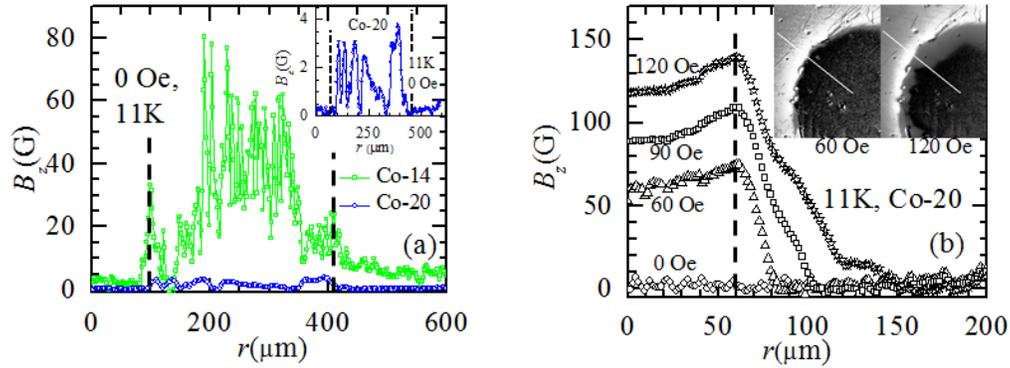

**Figure 5** (a) Distribution of local $B_z$ profile for Co-14 and Co-20 crystals at 0 Oe and 11 K. The dashed lines represent edges of the samples. The inset shows the zoomed-in image of the ($B_z$-$H$) distribution at 0 Oe and 11 K for Co-20 showing a very weak but finite positive magnetic response. (b) Variation of local $B_z$ distribution for Co-20 along the line shown in the inset of (b) at 11 K and different field with H//c . The Inset of (b) shows isofield MO images for Co-20 crystal at 11 K with $H$=60 Oe (left panel) and $H$=120 Oe (right panel).

0 Oe, shows that Co-14 crystal exhibits a comparatively larger positive $B_z$ response than that of Co-20 (c.f. Fig. 5a). Although small and much weaker than Co-14 crystal, the $B_z$ distribution deep inside Co-20 sample is not zero but finite with $B_z$~2 G (c.f. Inset Fig. 5a). This is consistent with the observation in Figure 2a-c that the paramagnetic features in bulk $M(T)$ are present in Co-20 crystal although they are much weaker as compared to Co-14 crystal response. Figure 5b presents the $B_z(r)$ for Co-20 at 11 K and different applied $H$ measured along the line shown in insets of Figure 5b. Figure 5b shows that unlike Co-14 crystal, $B_z(r)$ profiles in Co-20 are similar to the Bean critical state profiles [37] associated with vortex penetration in conventional superconductors.

**Discussion:**

Figure 6a-c present the ($H$-$T$) phase diagram constructed out of the characteristic temperatures identified from features in bulk magnetization (c.f. Fig. 2d-f) for the three crystals. The bluish region in Figure 6a-c below $T_{mf}^{on}$ (c.f. Fig. 2d-f) line represents the regime where only magnetic fluctuation response is sustained and the positive magnetization response in this regime increases with decrease in $T$. The yellow shaded region indicates the onset of superconducting fluctuations at $T_{scf}^{on}$ (above $T_c$) coexisting with regions with short range magnetic order (magnetic fluctuations) between $T_p$ and $T_{scf}^{on}$ (c.f. Fig. 2d-f and 4f). In the yellow shaded region the response from magnetic fluctuations dominates over the superconducting fluctuations (for example c.f. Fig. 4d-f for Co-14). With decreasing $T$ towards $T_c$ the response from the dominant magnetic fluctuations grows and also the diamagnetic response from superconducting fluctuations. As a result in this regime below $T_{scf}^{on}$ the net $M$ deviates significantly from the Curie

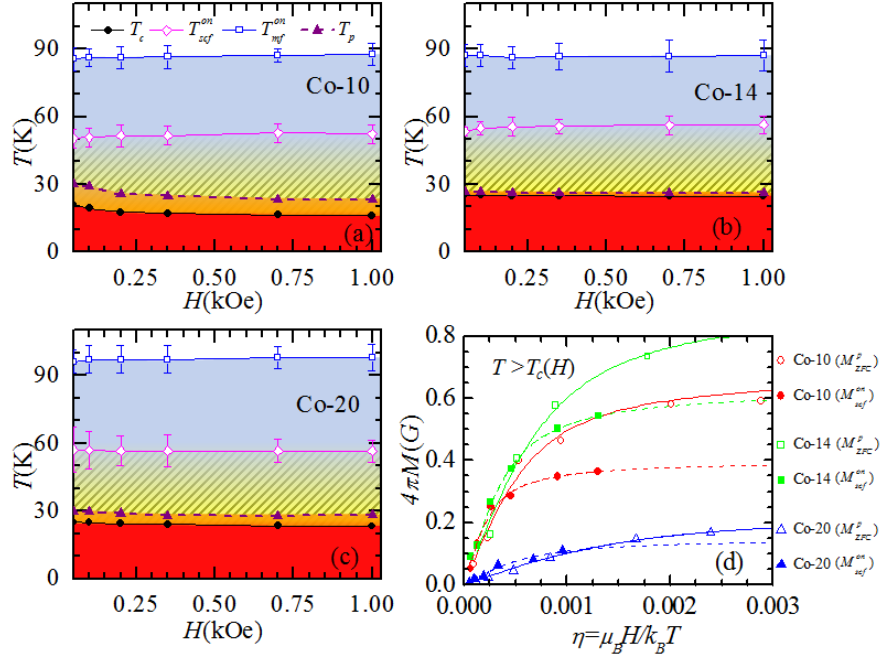

**Figure 6** Variation of $T_c$, $T_p$, $T_{mf}^{on}$ and $T_{scf}^{on}$ (c.f. Fig. 2a-f for identification) as a function of $H$ for (a) Co-10, (b) Co-14 and (c) Co-20. The error bars indicates the uncertainty in the estimated $T_{mf}^{on}$ and $T_{scf}^{on}$ values. (d) Variation of $4\pi M_{ZFC}^p$ (open symbols) and $4\pi M_{scf}^{on}$ (closed symbols) for all samples with $\eta = \mu_B H / k_B T$, where $\mu_B$ is the Bohr magneton. The solid and dashed lines represent fitting to the equation $M(\eta) = M_0 \left( Coth(a\eta) - 1/a\eta \right)$ with $4\pi M_{ZFC}^p$ and $4\pi M_{scf}^{on}$ respectively where $a = gJ\mu$ and $M_0$ is the saturated magnetization value of $4\pi M_{ZFC}^p$ and $4\pi M_{scf}^{on}$.

like behavior values expected if only magnetic fluctuations were present (cf. Fig. 2d-f). The actual $M$ data exhibits an peak feature at $T_p$ (as noted in Fig 2a-c and also in Fig. 4a) below which the diamagnetic response enhances at a faster rate causing $M$ to drop rapidly (c.f. Fig. 2a-c). In Figure 6a-c, the orange shaded region (between $T_p$ and $T_c$) is where superconducting fluctuations dominate over magnetic fluctuations causing a sharp change in $M$ from a positive value at $T_p$ towards a diamagnetic response below $T_c$. Finally below $T_c(H)$ bulk superconductivity sets in (red shaded region in Fig. 6a-c) with weak magnetic fluctuations surviving in the superconducting state as noted in Figure 3 and 5. The orange region (between $T_p$ and $T_c$ lines) is the narrowest for the OpD crystal (c.f. Fig. 6b) i.e., for Co-14 crystal the transition (from positive $M_{ZFC}^p$) into bulk superconducting state is the sharpest. The different lines in Figure 6a-c are weakly field dependent in the moderate field range however $T_c$ and $T_p$ lines are more sensitive to $x$ variation while $T_{mf}^{on}$ shifts only at higher $x = 0.2$ and $T_{scf}^{on}$ is almost independent of $x$.

Figure 6d shows the $4\pi M$ values determined from bulk $M(T)$ data (c.f. Fig. 2a-f) at $T = T_{scf}^{on}$ (viz., $M_{scf}^{on}$ values shown as closed symbol) and at $T = T_p$ ($M_{ZFC}^{p}$, open symbol) at different $H$, and plotted as a function of $\eta = \mu_B H / k_B T$ ($\mu_B$: Bohr magneton) for all three crystals. From the earlier discussion (c.f. Fig. 2d-f) $T \leq T_{scf}^{on}$ represents a regime where the $M$ is dominated by magnetic fluctuations. Here it appears that in spite of increase in doping concentration, the saturated value of $M_{scf}^{on}$ is highest in Co-14 crystal and weakest in Co-20. This suggests the observed feature of short range magnetic order (magnetic fluctuations) is not correlated with Co concentration. In Figure 6d the $M_{scf}^{on}$ and $M_{ZFC}^{p}$ are fitted to a Langevin like functional form, $M(\eta) = M_0 \left( Coth(a\eta) - 1/a\eta \right)$, where the fitting parameters are $M_0$ (the saturated magnetization value) and $a$. With local field $B=\mu H$ where $H$ is the externally applied magnetic field, we use the form $a\eta = gJ\mu_B(\mu H)/k_B T = gJ\mu_B B/k_B T$ in the Langevin expression to fit the data $M\left(\eta = \mu_B H/k_B T\right)$ (c.f. Fig. 6d), with $a = gJ\mu$ where $\mu$: the magnetic permeability, $g$ is the Lande $g$ factor and $J$ is the total angular momentum quantum number. We consider the quantity $gJ$ does not vary with doping while $\mu$ changes significantly. We determine values of the parameter $a$ by fitting $M_{scf}^{on}(\eta)$ and $M_{ZFC}^{p}(\eta)$ curves,

**Table 2**

| Crystal | $a_p$* | $a_{scf}^{on}$† | $v_{sf}$§ |
|---|---|---|---|
| | | | (%) |
| Co-10 (UD) | 3841 ± 473 | 8501 ± 945 | 0.094 |
| Co-14 (OpD) | 2748 ± 272 | 5198 ± 272 | 0.346 |
| Co-20 (OD) | 1445 ± 140 | 3615 ± 836 | 0.008 |

*From $M_{ZFC}^{p}(\eta)$ fitting.
†From $M_{scf}^{on}(\eta)$ fitting.
§Estimated from bulk ZFC $M(T)$ at $H$=50 Oe and $T=T_p$.

viz., $a_{scf}^{on}$ and $a_p$ which are listed in Table 2. A comparison of $a_{scf}^{on}$ and $a_p$ values for all the three crystals shows that $a_p$ is less than $a_{scf}^{on}$. At $T > T_{scf}^{on}$ the local $B$ experienced by moments in the crystal is higher due to the presence of regions with short range magnetic order. As $T$ decreases below $T_{scf}^{on}$ due to the appearance and subsequent enhancement in the strength of diamagnetic superconducting fluctuations, the effective $B$ decreases due to which $\mu$ reduces and hence lowering the value of $a_p$ in comparison to $a_{scf}^{on}$ value. This discussion also indicates that the presence of diamagnetic fluctuations in between $T_p$ and $T_{scf}^{on}$.

To estimate the superconducting volume fraction we define a quantity $v_{sf} = \delta M / 4\pi M (H=50\ Oe, T \ll T_c)$ at $T_p (>T_c)$ where $\delta M = 4\pi M_{ZFC}^{p} - 4\pi M_{ZFC,ex}^{p}$ with $4\pi M_{ZFC}^{p}$ the experimentally measured $M$ value at $T=T_p$ with $H$=50 Oe and $4\pi M_{ZFC,ex}^{p}$ is the corresponding extrapolated value of $M$ at $T_p$ (obtained from the linear Curie fitting in Fig. 2d-f). A schematic representation of $\delta M$ is shown in Figure. 2d. The denominator in $v_{sf}$ is the bulk superconducting response determined from the saturated diamagnetic value of $4\pi M$ at $H = 50$ Oe at $T \ll T_c$ (c.f. Fig. 1a-c). Table 2 shows the estimated $v_{sf}$ of 0.094% for Co-10 (UD), 0.346% for Co-14 (OpD) and 0.008% for Co-20 (OD) crystal. The largest superconducting fraction above $T_c$ is found for the Co-14 (OpD), which also has significant magnetic fluctuation (in fact it has the highest saturated value of $4\pi M_{scf}^{on}$, Fig. 6d) and a significant value of $a_{scf}^{on}$ ($\propto \mu$). We believe the comparatively larger $v_{sf}$ in Co-14 (OpD) crystal promotes long range superconducting phase coherence to develop in a smaller $T$ window between $T_p$ and $T_c$. Note that an image like Figure 4e only represents the response measured on the surface of the crystal at a particular location and it cannot be used to obtain an average information like the superconducting volume fraction. For this one would need to rely on bulk measurements alone. We speculate that regions with local diamagnetic response or superconducting fluctuations above $T_c$ as imaged in Figure 4d-f, could be similar to the Pseudo gap state found in high $T_c$ superconductors [45]. Above $T_c$ the presence of short range magnetic order (magnetic fluctuations) perhaps help in mediating superconducting pairing [21-28] resulting in localized regions with superconducting fluctuations with no macroscopic phase coherence. The close relationship between magnetic fluctuations and superconductivity is seen from observations in Co-14 crystal, which shows the strongest effect of magnetic fluctuations (c.f. largest $a_{scf}^{on}$ and $M_{scf}^{on}$ values) and it also has the largest $v_{sf}$ and the highest $T_c$ as well. In fact with weakening magnetic fluctuations as in Co-20 the superconducting fraction and $T_c$ also decreases significantly. The above suggests the possibility of magnetic fluctuations play an important role in mediating superconducting pairing to exist above $T_c$ in this compound.

Recently studies in FeSe crystals have shown evidence for BCS-BEC crossover in a Fermi system with strong spin imbalance, where the presence of preformed pairs in the crossover regime [32] was reported. The work suggested the possibility that in such systems with spin imbalance, the superconducting state maybe exotic with high degree of spin polarization, viz., spin triplet pairing. We note that above $T_c$ in our crystals of $BaFe_{2-x}Co_xAs_2$ we also observe evidence of the presence of superconducting fluctuations exhibiting local diamagnetic response embedded in a significantly magnetized background. However despite the presence of a positive magnetic response, we believe below $T_c$ the energy reduction in $BaFe_{2-x}Co_xAs_2$ crystals due to onset of superconductivity is greater than any increase in energy due to Zeeman contributions from any magnetized state surviving below $T_c$. In the $BaFe_{2-x}Co_xAs_2$ as the same set of electrons need to choose between participating in pairing or contributing to magnetism near $T_c$, the energy considerations favour pairing via magnetic fluctuations perhaps via magnonic states [13,46,47]. This may partially explain the predominance of the diamagnetic response as $H$ is increased (c.f. Figure 3f). Our observation also seems to suggests the absence of any exotic spin triplet pairing possibility in $BaFe_{2-x}Co_xAs_2$.

In conclusion we have presented evidence for the presence of superconducting fluctuations above $T_c$ in $BaFe_{2-x}Co_xAs_2$ with different Co - doping concentration, which coexists with regions with short range magnetic order. The optimally doped crystal with highest $T_c$ exhibits the strongest effect of magnetic fluctuations and it also possesses the highest volume fraction with superconducting fluctuations above $T_c$. Below $T_c$ along with bulk superconductivity weak magnetic fluctuations are sustained which are suppressed with the application of a magnetic field. The study suggests that in $BaFe_{2-x}Co_xAs_2$ magnetic fluctuations perhaps play an important role in generating local superconducting pairing correlations above $T_c$ which inturn helps for onset of bulk superconductivity in such systems. We hope our work motivates further investigations focused in this direction.

Supplementary:

**Magneto-Optical imaging (MOI).** The MOI technique is based on the principle of Faraday Effect, associated with the phenomenon of rotation of the plane of polarization of light by an angle proportional to the strength of the local field in the direction of propagation. Using a high sensitivity CCD camera, we image the Faraday rotated linearly polarized light (wavelength of 550 nm) intensity ($I(x,y)$, where ($x$, $y$) are the position co-ordinates on the surface being imaged). On the surface of $BaFe_{2-x}Co_xAs_2$ single crystals we place a Faraday active YIG thin film which has a reflecting layer in direct contact with the flat, freshly cleaved crystal surface. The YIG film senses the local field distributed on the surface of the crystal. The linearly polarized light which passes through the YIG film and gets reflected from the surface of the crystal undergoes a Faraday rotation due to the local field experienced by the YIG film. Expression for the Faraday rotation angle $\theta$ is given by $\theta = VB_z 2d$ where $V$ is wavelength dependent Verdet constant (maximum for 500 nm), $B_z$ is the component of the magnetic field along the wave vector of the light and $2d$ is the distance which the light wave traverses in the medium ($d$ being the thickness of the medium and the factor 2 for the case of reflection type Faraday rotation). From the Malus law, the intensity of the Faraday rotated light propagating in the direction of $B_z$ is $I(B_z) = I_0 Sin^2(2VB_z d) \propto B_z^2(x, y)$ for small $\theta$ or moderately low $B_z$. By sensitively imaging and calibrating the reflected Faraday rotated light intensity distribution ($I(x,y)$), one measures the local $B_z(x,y)$ across the surface of the sample, as $I(x,y) \propto [B_z(x,y)]^2$.

**MOI characteristics of a conventional superconductor.** Typically in a MO image for a conventional superconductor cooled below $T_c$ in $H = 0$ Oe there will be no contrast difference between regions inside the superconductor and outside it as there will be no Faraday rotation from inside the superconductor as well as outside since $B = 0$ G inside as well as outside. At a low non zero $H$ due to the Meissner effect as magnetic flux is expelled from inside the diamagnetic superconductor (where $B_z = 0$ G) and it is non zero outside the superconductor, one observes a dark image contrast over the superconductor while the contrast will be gray outside the superconductor where $B \neq 0$ as $H \neq 0$.

**Differential Magneto-optical Imaging (DMO).** In DMO technique we obtain differential images by periodically modulate the $H$ by an amount $\delta H = 1$ Oe and capturing $k$ nos of Faraday rotated images at $H$ and $H+\delta H$, viz., $I_i(H)$ and $I_i(H+\delta H)$, where $i$ is the index of the $k$ nos. of image captured. The differential image map, is $\delta I(x,y) =$ $\frac{\sum_{i=1}^{k} I_i(H+\delta H)}{k} - \frac{\sum_{i=1}^{k} I_i(H)}{k}$ ($k = 20$ is used). Using the above Malus law it can be shown that in response to external $H$ modulation, as $B_z$ changes by $\delta B_z$, $\delta I(x, y) \propto B_z \delta B_z$ for small rotation angles. Hence $\delta I/I \sim \delta B_z/B_z$ for moderately $B_z$ values.

**Acknowledgement**: SSB would like to acknowledge funding support from IIT Kanpur and DST and access to SQUID – VSM facility in TIFR, Mumbai for some of the measurements.

1. Ni, N. *et. al.* Effects of Co substitution on thermodynamic and transport properties and anisotropic $H_{c2}$ in Ba(Fe$_{1-x}$Co$_x$)$_2$As$_2$ single crystals. *Phys. Rev. B* **78**, 214515 (2008).
2. Lester C. *et al.* Neutron scattering study of the interplay between structure and magnetism in Ba(Fe$_{1-x}$Co$_x$)$_2$As$_2$. *Phys. Rev. B* **79**, 144523 (2009).
3. de la Cruz, C. *et al*. Magnetic order close to superconductivity in the iron-based layered LaO$_{1-x}$F$_x$FeAs systems. *Nature* **453**, 899-902 (2008).
4. Zhao, J. *et al.* Structural and magnetic phase diagram of CeFeAsO$_{1-x}$F$_x$ and its relation to high-temperature superconductivity. *Nature Mater.* **7**, 953-959 (2008).
5. Lu, D. H. *et al.* Electronic structure of the iron-based superconductor LaOFeP. *Nature* **455**, 81-84 (2008).
6. Liu, C. *et al.* K-doping dependence of the fermi surface of the Iron-Arsenic Ba$_{1-x}$K$_x$Fe$_2$As$_2$ superconductor using angle-resolved photoemission spectroscopy. *Phys. Rev. Lett.* **101**, 177005 (2008).
7. Singh, D. J. Properties of KCo$_2$As$_2$ and alloys with Fe and Ru: Density functional calculations. *Phys. Rev. B* **79**, 174520 (2009).
8. Avci, S. *et al.* Magnetoelastic coupling in the phase diagram of Ba$_{1-x}$K$_x$Fe$_2$As$_2$ as seen via Neutron diffraction. *Phys. Rev. B* **83**, 172503 (2011).
9. Williams, T. J. *et. al.* Magnetic and Superconducting Properties of Underdoped Ba(Fe$_{1-x}$Co$_x$)$_2$As$_2$ Measured with Muon Spin Relaxation/Rotation. arXiv:1408.3643 (2014).
10. Marsik, P. *et al.* Coexistence and Competition of Magnetism and Superconductivity on the Nanometer Scale In Underdoped BaFe$_{1.89}$Co$_{0.11}$As$_2$. Phys. Rev. Lett. **105**, 057001 (2011).
11. Pratt, D. K. *et al.* Incommensurate Spin-Density Wave Order in Electron-Doped BaFe$_2$As$_2$ Superconductors. Phys. Rev. Lett. **106**, 257001 (2011).
12. Williams ,T. J. *et. al.* Superfluid density and field-induced magnetism in Ba(Fe$_{1-x}$Co$_x$)$_2$As$_2$ and Sr(Fe$_{1-x}$Co$_x$)$_2$As$_2$ measured with muon spin relaxation. Phys. Rev. B 82, 094512 (2010).
13. Liu, R.H. *et al*. Large iron Isotope Effect in SmFeAsO$_{1-x}$F$_x$ and Ba$_{1-x}$K$_x$Fe$_2$As$_2$.Nature (London) **459**, 64 (2009).
14. Ryan, D. H. *et al.* Coexistence of long-ranged magnetic order and superconductivity in the pnictide Superconductor SmFeAsO$_{1-x}$F$_x$(x=0, 0.15). *Phys. Rev. B* **80**, 220503 (2009).
15. Laplace, Y. *et al.* Atomic coexistence of superconductivity and incommensurate magnetic order in the pnictide Ba(Fe$_{1-x}$Co$_x$)$_2$As$_2$. *Phys. Rev. B* **80**, 140501(R) (2009).
16. Drew, A. J. *et al.* Coexistence of static magnetism and superconductivity in SmFeAsO$_{1-x}$F$_x$ as revealed by muon spin rotation. *Nature Matre.* **8**, 310-314 (2009).
17. Julien, M. H. *et al.* Homogeneous vs. inhomogeneous coexistence of magnetic order and superconductivity probed by NMR in Co- and K-doped iron pnictides. *EPL* **87**, 37001 (2009).
18. Chu, J. H. *et. al.* Determination of the phase diagram of the electron-doped superconductor Ba(Fe$_{1-x}$Co$_x$)$_2$As$_2$. *Phys. Rev. B* **79**, 014506 (2009).
19. Chen, H. *et al.* Coexistence of the spin-density wave and superconductivity in Ba$_{1-x}$K$_x$Fe$_2$As$_2$. *EPL* **85**, 17006 (2009).
20. Mandal, P. *et al.* Anomalous local magnetic field distribution and strong pinning in CaFe$_{1.94}$Co$_{0.06}$As$_2$ single crystals. *EPL* **100**, 47002 (2012).
21. Cao, C. *et al.* Proximity of antiferromagnetism and superconductivity in LaFeAsO1−xFx: Effective Hamiltonian

from ab initio studies. *Phys. Rev. B* **77**, 220506(R)(2008).
22. Hirschfeld, P. J., Korshunov, M. M. & Mazin, I. I. Gap symmetry and structure of Fe-based superconductors. Rep. Prog. Phys. **74,** 124508 (2011).
23. Dai, X. *et al.* Even parity, orbital singlet, and spin triplet pairing for superconducting LaFeAsO$_{1-x}$F$_x$. *Phys. Rev. Lett.* **101**, 057008(2008).
24. Chubukov, A. V. Pairing mechanism in Fe-based superconductors. Annu.Rev.Condens. Matter Phys. **3,** 57-92 (2012).
25. Song, Can-Li. *et al.* Direct observation of nodes and twofold symmetry in FeSe superconductor. *Science* **332**, 1410 (2011).
26. Singh, D. J. & Du, M. H. Density functional study of LaFeAsO$_{1-x}$F$_x$: a low carrier density superconductor near itinerant magnetism. *Phys. Rev. Lett.* **100**, 237003 (2008).
27. Mazin, I. I., Singh, D. J., Johannes, M. D. & Du, M. H. Unconventional Superconductivity with a Sign Reversal in the Order Parameter of LaFeAsO$_{1-x}$F$_x$. *Phys. Rev. Lett.* **101**, 057003 (2008).
28. Kuroki, K. *et al.* Unconventional Pairing Originating from the Disconnected Fermi Surfaces of Superconducting LaFeAsO$_{1-x}$F$_x$. *Phys. Rev. Lett.* **101**, 087004 (2008).
29. Fernandes, R. M., Chubukov, A. V. & Schmalian J. What drives nematic order in iron-based superconductors? *Nature Physics* **10**, 97 (2014).
30. Shimojima, T. *et al.* Pseudogap formation above the superconducting dome in iron pnictides. Phys. Rev. B **89**, 045101 (2014).
31. Massee, F. *et al.* Pseudogap-less high T$_c$ superconductivity in BaCo$_x$Fe$_{2-x}$As$_2$, Euro. Phys. Lett. **92**, 57012 (2010).
32. Kasahara Shigeru *et. al*. Field-induced superconducting phase of FeSe in the BCS-BEC cross-over. *Proc. Natl. Acad. Sci*. USA **111**, 16309 (2014).
33. Vinod, K., Satya, A. T., Sharma, S., Sundar, C. S., & Bharathi, A. Upper critical field anisotropy in BaFe$_{2-x}$Co$_x$As$_2$ single crystals synthesized without flux. *Phys. Rev. B*.**84**, 012502 (2011).
34. Johansen, T.H. *Magneto-Optical imaging, Nato Sci. Ser. II* **142** (2004).
35. Wijngaarden, R. J. *et.al.* Fast imaging polarimeter for magneto-optical investigations. *Rev. Sci. Instrum.* **72**, 2661 (2001).
36. Mandal, P., Chowdhury, D., Banerjee, S. S. & Tamegai,T. High sensitivity differential magneto-optical imaging with a compact Faraday modulator. *Rev. Sci. Instrum.* **83**, 123906 (2012).
37. Bean, C. P. Magnetization of high field superconductors. *Rev. Mod. Phys.* **36**, 31(1964).
38. Nakajima, Y. *et. al.* Enhancement of critical current density in Co-doped BaFe$_2$As$_2$ with columnar defects introduced by heavy-ion irradiation. *Phys. Rev. B* **80**, 012510 (2009).
39. Koshelev A. E. & Larkin A. I. Paramagnetic moment in field-cooled superconducting plates: Paramagnetic Meissner effect. *Phys. Rev. B* **52**, 13559 (1995).)
40. Das, P. *et. al.* Surface superconductivity, positive field cooled magnetization, and peak-effect phenomenon observed in a spherical single crystal of niobium. *Phys. Rev. B* **78**, 214504 (2008).
41. Soibel, A. *et. al.* Imaging the vortex-lattice melting process in the presence of disorder. *Nature* **406,** 282–287 (2000).
42. Banerjee, S. S. *et al.* Melting of ''Porous'' Vortex Matter. *Phys. Rev. Lett.* **90**, 087004 (2003).
43. Shaw, G., Mandal, P., Banerjee S S & Tamegai, T. Visualizing a dilute vortex liquid to solid phase transition in a Bi$_2$Sr$_2$CaCuO$_8$ single crystal. *New J. Phys.* **14**, 083042 (2012).
44. Kalisky, B. *et. al.* Stripes of increased diamagnetic susceptibility in underdoped superconducting Ba(Fe$_{1-x}$Co$_x$)$_2$As$_2$ single crystals: Evidence for an enhanced superfluid density at twin boundaries. *Phys. Rev. B* **81**, 184513 (2010).
45. Pasupathy, A. N. *et. al.* Electronic Origin of the Inhomogeneous Pairing Interaction in the High-$T_c$ Superconductor Bi$_2$Sr$_2$Ca$_2$CuO$_{8+x}$. *Science* **320**, 196 (2008).
46. Zhou Jin - Ke et. al. Persistent high-energy spin excitations in iron-pnictide superconductors, Nature Commn.4, 1470 (2013).
47. Chen C.-C., Jia C. J., Kemper A. F., Singh R. R. P., and Devereaux T. P. Theory of Two-Magnon Raman Scattering in Iron Pnictides and Chalcogenides. *Phys. Rev. Lett.* **106**, 067002.